# Real time data access log analysis system of EAST tokamak based on spark

F. Wang[1], *Member IEEE*, Q.H. Zhang[2], X.Y. Sun[1], Y. Chen[1], Y.T. Wang[2], F. Yang[3]

*Abstract*—The experiment data generated by the EAST device is getting larger and larger, and it is necessary to monitor the MDSplus data storage server on EAST. In order to facilitate the management of users on the MDSplus server, a real-time monitoring log analysis system is needed. The data processing framework adopted by this log analysis system is the Spark Streaming framework in Spark ecosphere, whose real-time streaming data is derived from MDSplus logs. The framework also makes use of key technologies such as log monitoring, aggregation and distribution with framework likes Flume and Kafka which makes it possible for MDSplus mass log data processing power. The system can process tens of millions of unprocessed MDSplus log information at a second level, then model the log information and display it on the web. This report introduces the design and implementation of the overall architecture of real time data access log analysis system based on spark. Experimental results show that the system is proved to be with steady and reliable performance and has an important application value to the management of fusion experiment data. The system has been designed and will be adopted in the next campaign and the system details will be given in the paper.

## I. Introduction

The Experimental Advanced Superconducting Tokamak (EAST) is a lager fusion research device which has produced mass experimental raw data [1-3]. The high-volume database such as MDSPlus database which is a set of software tools for data acquisition and storage and a methodology for management of complex scientific data has stored more than five hundred TB experimental raw data that includes diagnostic DAQ raw data, analyzed data and engineering DAQ raw data, etc. [4]. So it is important for manager to watch the information and status of all the MDSplus data.

At present all data access behavior cannot be detailed recorded on MDSPlus logs except poor EAST data access logs which are stored into mdsipd file. The whole log information including client's link information and data operation information are not fully recorded. Only the simple information is recorded as list below.
- Client's TCP/IP connection;
- Client's TCP/IP disconnection;
- Valid information;

It is really hard for manager to monitor the storage system based on above information. Problems will be occur when some hackers attack the server. In addition, the pressure can be formatted when a lot of clients access a single node of the server. In this case, the problem causes the traffic congestion on the storage server. To develop a real-time data access log system to watch all data status became much more significant. All functions of the real time data access log analysis system is made up of four components as list below.
- Real-time data status monitoring;
- Real-time client operation monitoring;
- Off-line data analysis;
- Data browser;

These functions aim at solving the problem of missing monitor on storage server. Real-time data status monitoring can help the administrator to make precise decisions, and then carry out effective operation. According to the access log analysis system, administrator can pinpoint the IP addresses of people who have malicious access to the database, and then restrict its address access to prevent the server from crashing.

Beyond that, the off-line data analysis is also designed for data mining. Manager can find the most accessed shot number and the most logged in IP, just to name a few. Other key features, data browser visualize the real-time analysis data through the web charts.

The whole system is based on big data technology especially the key technology, spark that provide the spark streaming components for real-time batch data processing. We intend to construct a real-time data access log system of EAST Tokamak based spark that contains the above functions.

## II. System Architecture

According to the existing requirements analysis and current data log system, a new real-time data access log analysis system architecture has been designed as shown in Fig.1. The whole log analysis system consists of 5 parts.

F. Wang is with the Institute of Plasma Physics, Chinese Academy of Sciences, 350 Shushanhu Road, Hefei, Anhui 230031, P. R. China (phone: 86-551-65592311; e-mail: wangfeng@ipp.ac.cn).

X. Y. Sun, Y. Chen are both with the Institute of Plasma Physics, Chinese Academy of Sciences, 350 Shushanhu Road, Hefei, Anhui 230031, P. R. China (e-mail: xysun@ipp.ac.cn, cheny@ipp.ac.cn).

Q.H Zhang, T. Y. Wang are both with the Institute of Plasma Physics, Chinese Academy of Sciences, 350 Shushanhu Road, Hefei, Anhui 230031, P. R. China and University of Science and Technology of China, Hefei, Anhui 230026, P. R. China.

F. Yang is with the Department of Computer Science, Anhui Medical University, Hefei, Anhui 230032, P. R. China (email: fyang@ipp.ac.cn).

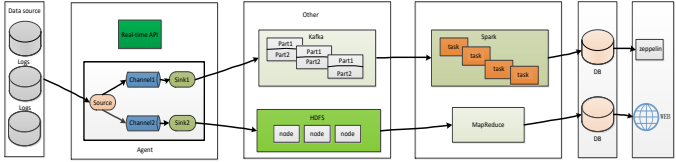

Fig. 1. System architecture

## A. Part1

Part1 is the improvement of MDSplus logging system, which can record the detailed log information when remote client sends requests to the server. This part plays an important role in the whole system because it's the data source. But it's worth noting that the logs just provide the raw data without any processing. The implement of the logging system makes full use of the hook functions. The detailed information about part1 is given in the next chapter.

## B. Part2

Part2 is the logging data monitor, which uses the open source flume logging framework and other real-time API. To respond to changes in log information in real-time, the flume server daemon watches the server log's file content all the time. When the data is collected, the flume sends the data to different destinations, both HDFS (Hadoop distributed file system) and Kafka (topics subscription and release system, converting log messages into streaming data) [5]. Flume was chosen as a logging monitor because it's reliable, fault-tolerant, scalable, manageable, and customizable.

## C. Part3

Part3 are both the data backup storage designed based on HDFS storage and the streaming data conversion based on Kafka. On the one hand, using HDFS storage to prepare for off-line processing. On the other hand, Kafka make log messages became streaming data which is data source of spark streaming procedure.

## D. Part4

Part4 is the streaming data processing program based on spark streaming. Each spark steaming Job has been divided into many parallel tasks. Each task can process one batch data from Kafka server. In this case, it acts as the consumer role. At the same time, upstream Kafka became producer role. In addition to real-time part, the MR calculation can perform offline data mining.

## E. Part5

Part5 is the data browser section including web presentation and Zeppelin which is a big data visualization tools [6]. The web presentation adapts traditional technology such as JS, JQuery, Echart, etc. The Zeppelin provides interfaces for many applications whether Hadoop or Spark or traditional relational databases.

## III. SYSTEM DESIGN

At present, access log analysis system is the only way to monitor the Tokamak experimental data so the real-time characteristic of system is very important for administrator's decision. Although there are already a lot of very sophisticated log management systems on the market, but these systems do not meet the specific log data fusion experiments. Spark Streaming is chosen instead of Strom (Strom is another instant framework) owing to the experimental data operation is a batch operation. This characteristic is much closer to Spark Streaming's data processing features. In another respect, Spark ecosphere can be easily deployed on existing Hadoop clusters. So we choose spark as the real-time processing engine [7-8].

### A. Log optimization

Log is optimized through Hook Function. We build our own log DDL. The log system has been optimized as shown in Fig.2.

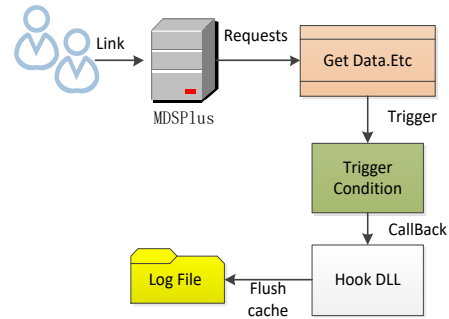

Fig. 2. Log system flow diagram

From the Fig.2 we can see the log system has been greatly improved. The whole process is shown as list below.
- Thin client mode connects MDSPlus server.
- Operations such as Tree Open, Get Data, Put Data will trigger TreeCallHook function and be recorded.
- libTreeShrHook.so flush the log information into mdsip log file
- Client disconnects link.

### B. Real-time log Collection

The MDSplus log message is sent to HDFS and Kafka through the monitoring of the Flume. As shown in Fig.3, the flume supports multiplexing the event flow to one or more destinations.

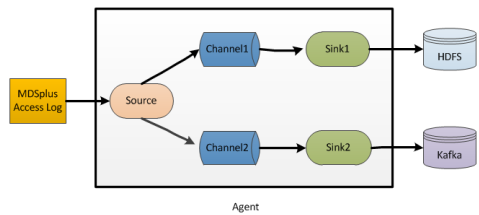

Fig. 3. Log system flow diagram

On the one hand, the log message through channel1 (disk cache) is sent to HDFS for storage, on the other hand, it is sent to Kafka for converting through channel2 (memory buffer).

Once Kafka server receives message from Flume server, it can put message into topic, and then Spark streaming pull message from the topic. The reason for not using the direct connection between flume and spark is that if the server produces so much message that processing server cannot handle it. In this case, that may cause procedure crash. There are many different versions about flume and Kafka. We design the real-time log collection system as flowing configuration.

- Flume NG 1.7.0.
- Kafka 2.11.
- Linux version 2.6.32-696.el6.x86_64.

### C. Real-time log analysis

The Spark Streaming can analysis the batch real-time data streaming which is pulled from Kafka topics and can be described as RDD (resilient distributed datasets). By a series of transformation, the RDD can be acted as the final format that we wanted. Event some mistakes occur at RDD's transformation, the RDD can be ability restored from parent's RDD which's called lineage [9]. Such RDD constituted the DAG. One of the real-time processing jobs is showed as DAG on Fig.4.

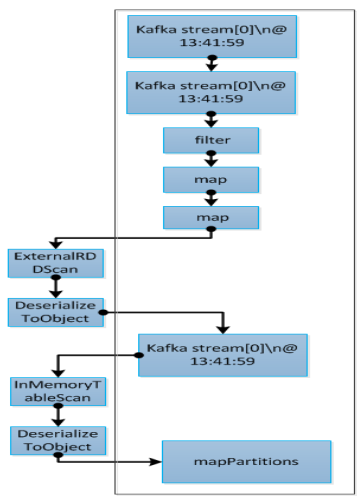

Fig. 4. Data flow schema

As you can see, the job is divided into different stages. At the same time, the stage is divided into different tasks. While the RDD's transformation includes filter, map, and so on. At last the operation of action write the final result into MYSQL. Two main format of the log message after processing in real time as Fig.5 and Fig.6 shown below.

Fig. 5. Client information

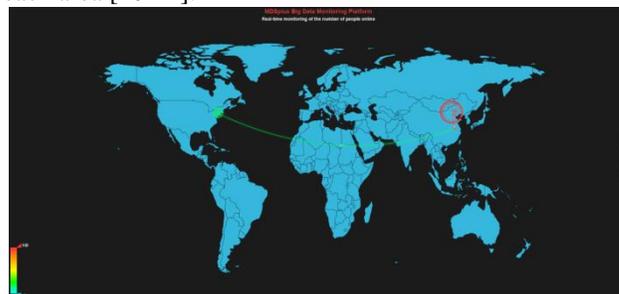

Fig. 6. Operation information

The client table does record five kinds segment including linkTime, pid, user, host and status while the operation table records six kinds segments including linkTime, pid, hooktype, tree, shot and nodepath.

### D. Data browser

The MySQL table does not directly reflect the value of the data. To solve this problem, building a data browser is quite necessary. Combining Zeppelin with traditional web can present server status perfectly. Fig.5. is one of log data visualization content which can show numbers of online client in each area [10-11].

Fig. 7. Web show server status

## IV. TEST RESULTS

To test the log analysis system's usability, the test method adapts multiply threads access data storage server. The following Table I is an off-line and real-time comparison of the log information processing.

TABLE I. HANDLING PERFORMANCE TEST

| Test Case | Speed pieces/s |
|---|---|
| Real-time | ~1,000,000 |
| Off line | ~3,000 |

And through the spark job monitor, one of the application's jobs just spends about 27ms in processing about 20,000 pieces of log information.

The following is the testing environment.
- Intel E5-2650 v2 @ 2.60GHz CPU / 64GB RAM
- Intel DC P3700 NVMe SSD.
- CentOS 6.8 64bit
- MySQL 5.1 / PHP 5.3.3 / Apache 2.2.15.

## V. Summary

To monitor the MDSPlus data storage server on EAST, a new real-time access log analysis system has been designed which includes 5 parts including log optimization, real-time log information collection, log information storage, real-time log analysis and data browser. The log metadata can flow between part1 and part2/part3/par4/part5 controlled by the system schedule program. The real-time data access log analysis system has been implemented and adopted in the next campaign of East tokamak.

In the future more real-time analysis components will be added into the log analysis system to mining more useful data, and more advanced machine learning algorithm will be implemented according to the requirements.

## Acknowledgment

This work is supported by the National Key R&D Program of China (Grant No: 2017YFE0300500, 2017YFE0300505). We would like to thank all the EAST Team members for their support.

## References

[1] B.J.Xiao, Q.P.Yuan, et al., "Recent plasma control progress on EAST", Fusion Engineering and Design, Vol. 87, p1887-1890, 2012.
[2] Liu Ying, Li Guiming, Zhu Yingfei, Li Shi, "New developments of the EAST data system", Fusion Engineering and Design Vol. 86, p151-154, 2011.
[3] Stillerman J A, Fredian T W, Klare K A, et al. MDSplus data acquisition system[J]. Review of Scientific Instruments, 68(1):939-942,1997
[4] MDSplus, http://www.mdsplus.org
[5] Kafka, http://kafka.apache.org
[6] Zeppelin,http://zeppelin.apache.org/
[7] F. Wang, Y. Chen, S. Li, Y. Wang, X.Y. Sun, F. Yang, "Study of Fast Data Access Based on Hierarchical Storage for EAST Tokamak", 2016 IEEE-NPSS Real Time Conference (RT), 2016.
[8] F. Wang, S. Li, Y. Chen, F. Yang, "The design of data storage system based on Lustre for EAST", Fusion Engineering and Design, April, 2016.
[9] Hunter T,Das T,Zaharia M,et al. Large-scale online expectation maximization with spark streaming,2012
[10] F. Yang, B. J. Xiao, "A web based MDSplus data analysis and visualization system for EAST", Fusion Engineering and Design, Volume 87, Issue 12, p2161-2165, 2012.
[11] Wang Z, Zhang C. Design and implementation of a data visualization analysis component based on ECharts[J]. Microcomputer & Its Applications, 2016.